\documentclass[onecolumn]{pasj02}
\usepackage{xcolor}

\jyear{2026}
\Received{}%{yyyy/mm/dd}
\Accepted{}%{yyyy/mm/dd}
\begin{document} 

\title{From supernovae to neutron stars: crust formation time}

%%% begin:list of authors
% Do NOT capitalize all letters in "textsc".

\author{Yudai \textsc{Suwa}\altaffilmark{1,2}}%
\email{suwa@g.ecc.u-tokyo.ac.jp}

\author{Ken'ichiro \textsc{Nakazato}\altaffilmark{3}}%

\altaffiltext{1}{Department of Earth Science and Astronomy, The University of Tokyo, Tokyo 153-8902, Japan}
\altaffiltext{2}{Center for Gravitational Physics and Quantum Information, Yukawa Institute for Theoretical Physics, Kyoto University, Kyoto 606-8502, Japan}
\altaffiltext{3}{Faculty of Arts and Science, Kyushu University, Fukuoka 819-0395, Japan}

%%% end:list of authors

%% !!! Select 3 to 5 words from PASJ's key words !!! 
%% List of Key Words: https://academic.oup.com/pasj/pages/Pasj_Keywords 
%% "\KeyWords{ }" always has to be placed before ``\maketitle'' 

\maketitle

\begin{abstract}
A neutron star is born as a hot, lepton-rich protoneutron star (PNS) and cools via neutrino emission, eventually allowing heavy ions in the outer layers to crystallize into a solid crust. We develop a simple analytic estimate for the onset time of this crust formation during the late, post-convective PNS cooling phase. Using a diffusion-based neutrino luminosity and the resulting entropy evolution together with an approximately isentropic interior structure, we obtain the time-dependent density and temperature at the neutrinosphere. We then impose the Coulomb crystallization condition for heavy nuclei, expressed through the Coulomb coupling parameter, and determine when the neutrinosphere temperature first falls below the crystallization threshold evaluated at the neutrinosphere density. This procedure yields closed expressions for the entropy at crystallization and the corresponding crust-formation time, with explicit dependence on the PNS mass and radius, an effective diffusion/cooling normalization, and composition parameters such as the ionic charge $Z$ and heavy-nuclei mass fraction. For canonical microphysics, we find that the first solid phase typically appears at $t_{\mathrm{crust}}\sim 100$--$500\,\mathrm{s}$. These closed-form scalings provide a useful late-time analytic benchmark for the onset of crust formation and clarify its dependence on PNS and composition parameters.
\end{abstract}

%\pagewiselinenumbers 

\section{Introduction}\label{sec:introduction}

Neutron stars (NSs) are born in core-collapse supernovae as hot, lepton-rich protoneutron stars (PNSs). In the first seconds after bounce, a PNS has temperatures of $\sim 10^{10}$--$10^{11}\ \mathrm{K}$ and cools predominantly by emitting neutrinos, which carry away a substantial fraction of the gravitational binding energy ($\sim 10^{53}\ \mathrm{erg}$) on a timescale of $\sim 10$--$100\ \mathrm{s}$ \citep{1986ApJ...307..178B,1999ApJ...513..780P,2012ARNPS..62..407J,2013ApJS..205....2N,2019ApJ...881..139S}. This early cooling and deleptonization phase sets the initial thermodynamic structure of the nascent NS and provides the physical conditions for the emergence of a solid crust.

The formation of the crust is a key milestone in the transition from a fluid PNS to a ``cold'' NS. At sufficiently high temperature, ions behave as a strongly coupled plasma, and the stellar matter is effectively fluid-like in the sense that it cannot sustain elastic shear stresses. As the star cools, Coulomb interactions between ions eventually dominate over thermal agitation and drive crystallization into a Coulomb lattice \citep{shap83,2007ASSL..326.....H,2008LRR....11...10C}. Once an elastic crust forms, it qualitatively changes several aspects of NS physics: it enables the storage of elastic strain, affects the coupling between the superfluid/superconducting interior and the magnetized exterior, and influences the subsequent evolution of the magnetic field and rotation (see \citet{2019LRCA....5....3P} for a review). For newborn NSs, the emergence of a solid crust may also provide a useful transition time for later magnetic-field evolution.

Despite its importance, the crust-formation time is not a single universal value. It depends on the cooling history (governed mainly by neutrino transport and the PNS radius/mass evolution), and on the microphysics that determines the melting temperature of the ionic component (composition, charge number $Z$, and the Coulomb coupling at melting) \citep{1999ApJ...513..780P,2007ASSL..326.....H,2008LRR....11...10C}. Fully self-consistent numerical simulations with neutrino radiation transport can follow the PNS evolution in detail \citep{1999ApJ...513..780P,2010AandA...517A..80F,2010PhRvL.104y1101H,suwa14,naka18}, but they are computationally expensive and not always suited for scanning broad parameter dependencies. For many applications, an analytic or semi-analytic estimate with explicit scaling relations is valuable: it clarifies which macroscopic parameters control the result and allows rapid exploration across plausible PNS configurations.

In this paper, we present a simple analytic estimate for the crust-formation timescale during the late PNS cooling phase. Our approach combines (i) an analytic description of the neutrino luminosity and entropy evolution based on the diffusion approximation \citep{suwa21}, together with the approximately isentropic thermal structure found in the late cooling phase \citep{suwa14,naka18}, which we interpret as a simplified post-convective PNS profile,\footnote{ PNS convection enhances lepton and energy transport and can drive much of the PNS interior toward a nearly isentropic state within the first few seconds after bounce, with characteristic entropies of order $s\sim1$--$3\,k_{\rm B}$ per baryon depending on the EOS \citep{2012PhRvL.108f1103R,2016NCimR..39....1M,2020MNRAS.492.5764N,2022MNRAS.511..356P,2025ARNPS..75..425J,2026arXiv260209025R}. We do not model convection explicitly.} and (ii) the crystallization criterion for a Coulomb lattice expressed through the Coulomb coupling parameter \citep{shap83}. This yields the time-dependent trajectory of the PNS thermodynamic profile in the $\rho$--$T$ plane, which we compare to the critical temperature for crystallization. Adopting representative nuclear composition parameters motivated by cold-catalyzed matter calculations \citep{1971ApJ...170..299B} and allowing for plausible variations in $Z$, we derive an explicit estimate of the epoch at which the PNS first satisfies the crust-formation condition near the neutrinosphere/outer layers. The resulting crust-formation time is typically of order $\sim 10^2\ \mathrm{s}$ for canonical parameters, while exhibiting strong and explicit dependence on the PNS mass, radius, and the neutrino-cooling normalization.

This paper is organized as follows. In Section~\ref{sec:cooling}, we summarize the analytic description of neutrino cooling and the resulting entropy and temperature evolution. In Section~\ref{sec:crust}, we describe the crystallization condition and derive the crust-formation time by matching the PNS thermal trajectory to the critical line in the $\rho$--$T$ plane. 
In Section~\ref{sec:summary}, we summarize our results and discuss their interpretation and limitations.

\section{Neutrino cooling}\label{sec:cooling}

This section summarizes the analytic PNS cooling model that we use to obtain the time-dependent thermal trajectory in the $\rho$--$T$ plane. We focus on the quasi-stationary late cooling phase, after convective mixing has helped flatten the entropy profile, and use a diffusion-based parametrization for the subsequent thermal evolution \citep{suwa14,naka18}. For the purpose of a scaling estimate, we adopt the diffusion-based analytic solution developed by \citet{suwa21}. We note that the analytic formulae were calibrated against detailed cooling calculations mainly up to $\sim$\,several$\times 10\ {\rm s}$ in \citet{2019ApJ...881..139S}; in this work we use them as an extrapolative parametrization to later times in order to derive explicit scaling relations for crust formation.

The neutrino transport equation can be derived from the Boltzmann equation. State-of-the-art core-collapse supernova simulations solve the transport problem with minimal approximations (see, e.g., \cite{2012ARNPS..62..407J,2024PJAB..100..190Y} for reviews), but such calculations are computationally expensive for broad parameter surveys and long-term integrations. Here we therefore employ the diffusion approximation, which is appropriate in optically thick regions and provides simple analytic scaling relations. In the diffusion limit, the neutrino energy flux is written as
\begin{align}
F=-\frac{c}{3\kappa}\frac{\partial \epsilon_\nu}{\partial r},
\label{eq:diff_flux}
\end{align}
where $F$ is the diffusive neutrino flux, $c$ is the speed of light, $\kappa$ is the total (transport) opacity, and $\epsilon_\nu$ is the neutrino energy density. The outward neutrino luminosity $L$ is then
\begin{align}
L(r) &= 4\pi r^{2} F(r),
\label{eq:diff_L}
\end{align}
so that the luminosity at the neutrinosphere is $L=L(R_\nu)$. Because neutrinos decouple there, this $L$ is the outward luminosity seen by a distant observer. We use this diffusion-based relation as the starting point for the late-time analytic luminosity solution below.

Following \citet{suwa21}, the late-time analytic solution for the neutrino luminosity can be expressed as
\begin{align}
L = 3.3 \times 10^{51} \, \rm{erg\, s}^{-1}
\left(\frac{M_{\rm{PNS}}}{1.4 M_\odot}\right)^{6}
\left(\frac{R_{\rm{PNS}}}{10 \, \rm{km}}\right)^{-6}
\left(\frac{g\beta}{3}\right)^{4}
\left(\frac{t + t_0}{100\,\rm{s}}\right)^{-6},
\end{align}
where $M_{\rm PNS}$ and $R_{\rm PNS}$ are the PNS mass and radius, $t$ is the time variable in the analytic cooling solution, and $t_0$ sets the time origin (i.e., the initial condition of the entropy; see below). The dimensionless factor $g\beta$ is an effective parameter introduced by \citet{suwa21}, which combines (i) a density-profile correction factor $g$ (accounting for deviations of the near-surface density gradient from the $n=1$ Lane--Emden form) and (ii) an opacity-boosting factor $\beta$ (e.g., due to coherent scattering; $\beta$ may increase when heavy nuclei appear). In what follows, we treat $g\beta$ as a constant parameter controlling the diffusion timescale. Here $L$ denotes the luminosity per neutrino species in the convention of \citet{suwa21}; the total luminosity summed over $\nu$ and $\bar{\nu}$ flavors is therefore $L_{\rm tot}=6L$.

The parameter $t_0$ is determined by the total energy emitted in neutrinos. Integrating the luminosity over time gives
\begin{align}
t_0 = 210\, \rm{s}
\left(\frac{M_{\rm{PNS}}}{1.4 M_\odot}\right)^{6/5}
\left(\frac{R_{\rm{PNS}}}{10 \, \rm{km}}\right)^{-6/5}
\left(\frac{g\beta}{3}\right)^{4/5}
\left(\frac{E_{\rm tot}}{10^{52}\,\rm{erg}}\right)^{-1/5},
\end{align}
where $E_{\rm tot}$ is the total energy emitted by neutrinos.

Next, we derive the temperature distribution inside the PNS by assuming an approximately constant entropy profile, which is a good approximation in the late cooling phase \citep{suwa14,naka18}. In a degenerate Fermi gas, the entropy per baryon scales as
\begin{align}
s \sim \frac{k_{\rm B}T}{\epsilon_{\rm F}},
\end{align}
where $s$ is the entropy per baryon and $\epsilon_{\rm F}$ is the nucleon Fermi energy (this expression is accurate up to numerical factors of order unity; see, e.g., \cite{beth90}). Approximating $\epsilon_{\rm F}$ by the non-relativistic nucleon Fermi energy yields
\begin{align}
\epsilon_{\rm F}
=\frac{\hbar^{2}}{2m_{\rm N}}(3\pi^{2}n_{\rm B})^{2/3}
= 30.4\,{\rm MeV}
\left(\frac{\rho}{10^{14}\,{\rm g\,cm^{-3}}}\right)^{2/3},
\label{eq:epsilonF}
\end{align}
where $m_{\rm N}$ is the nucleon mass, $n_{\rm B}$ is the baryon number density ($n_{\rm B}= \rho/m_{\rm N}$), and $\rho$ is the mass density. The temperature profile then follows as
\begin{align}
T
\approx \frac{s\,\epsilon_{\rm F}}{k_{\rm B}}
= 6.5\times 10^{10}\,{\rm K}
\left(\frac{\rho}{10^{14}\,{\rm g\,cm^{-3}}}\right)^{2/3}
\left(\frac{s}{1\,k_{\rm B}/{\rm baryon}}\right).
\label{eq:T_profile}
\end{align}

To connect the interior profile to the outer boundary condition, we estimate the neutrinosphere position. We adopt the $n=1$ Lane--Emden density profile used in \citet{suwa21},
$\rho(\xi)=\rho_c \sin\xi/\xi$ with $r=\alpha\xi$, where $\alpha=R_{\rm PNS}/\pi$ and $\rho_c=M_{\rm PNS}/(4\pi^{2}\alpha^{3})$. The neutrinosphere radius is written as $R_\nu=\alpha\xi_\nu$ and is approximated by 
\begin{align}
R_\nu
=\alpha\xi_\nu
=R_{\rm PNS}
\left[
1-0.019
\left(\frac{M_{\rm PNS}}{1.4M_\odot}\right)^{-7/10}
\left(\frac{R_{\rm PNS}}{10\,{\rm km}}\right)^{9/5}
\left(\frac{g\beta}{3}\right)^{-3/10}
\left(\frac{s}{1\,k_{\rm B}/{\rm baryon}}\right)^{-3/5}
\right].
\end{align}
The corresponding density at the neutrinosphere is
\begin{align}
\rho_\nu
:=\rho(R_\nu)
=\rho_c \frac{\sin(R_\nu/\alpha)}{R_\nu/\alpha}
\approx \rho_c \frac{\pi-\xi_\nu}{\pi}.
\end{align}
Using the approximate expression for $\xi_\nu$ given in \citet{suwa21} (their Eq.\ 40), the neutrinosphere density can be estimated as
\begin{align}
\rho_\nu
= 4.1 \times 10^{13}\,{\rm g\,cm^{-3}}
\left(\frac{M_{\rm PNS}}{1.4M_\odot}\right)^{3/10}
\left(\frac{R_{\rm PNS}}{10\,{\rm km}}\right)^{-6/5}
\left(\frac{g\beta}{3}\right)^{-3/10}
\left(\frac{s}{1\,k_{\rm B}/{\rm baryon}}\right)^{-3/5}.
\label{eq:rho_nu}
\end{align}
Evaluating Eq.\ (\ref{eq:T_profile}) at $\rho=\rho_\nu$ gives the neutrinosphere temperature
\begin{align}
T_\nu
:=T(R_\nu)
= 3.6 \times 10^{10}\,{\rm K}
\left(\frac{M_{\rm PNS}}{1.4M_\odot}\right)^{1/5}
\left(\frac{R_{\rm PNS}}{10\,{\rm km}}\right)^{-4/5}
\left(\frac{g\beta}{3}\right)^{-1/5}
\left(\frac{s}{1\,k_{\rm B}/{\rm baryon}}\right)^{3/5}.
\end{align}
Outside the neutrinosphere, an atmosphere forms and can be approximated as nearly isothermal with $T=T_\nu$ in the late cooling phase.

Finally, the time evolution of the entropy is obtained from the energy balance between the PNS thermal energy and neutrino emission. In the analytic model of \citet{suwa21}, combining $dE_{\rm th}/dt=-6L$ (accounting for all neutrino/antineutrino flavors) with the Lane--Emden structure yields
\begin{align}
s
= 4.0\, k_{\rm B}/{\rm baryon}
\left(\frac{M_{\rm PNS}}{1.4M_\odot}\right)^{13/6}
\left(\frac{R_{\rm PNS}}{10\,{\rm km}}\right)^{-2}
\left(\frac{g\beta}{3}\right)^{2}
\left(\frac{t+t_0}{100\,{\rm s}}\right)^{-5/2}.
\label{eq:s_of_t}
\end{align}
Combining Eqs.\ (\ref{eq:T_profile}), (\ref{eq:rho_nu}), and (\ref{eq:s_of_t}) gives the time-dependent thermodynamic trajectory of the PNS in the $\rho$--$T$ plane (and, in particular, $\rho_\nu(t)$ and $T_\nu(t)$), which we will compare with the crystallization condition in the next section.

Before proceeding, we emphasize the applicability range of this analytic treatment. The cooling formulae were calibrated mainly up to times of order several$\times10\,\mathrm{s}$ in detailed models \citep{2019ApJ...881..139S}, whereas in this work they are extrapolated to later epochs to obtain explicit crust-formation scalings. Detailed PNS cooling calculations also indicate that at epochs of order $t\sim100\,\mathrm{s}$ and later the interior becomes increasingly semitransparent to neutrinos \citep{1999ApJ...513..780P,2010AandA...517A..80F,2010PhRvL.104y1101H,2022MNRAS.511..356P}. At these later times, neutrino transport is therefore no longer everywhere well described by a purely diffusive picture. Accordingly, the crust-formation time, $t_{\rm crust}$, derived below should be regarded as useful late-time analytic benchmarks for the onset of crust formation, intended to clarify the parameter dependences rather than to provide model-specific predictions for individual PNSs.

% ---------------------------------------
\section{Crust formation time}\label{sec:crust}

In this paper we define the crust-formation time $t_{\rm crust}$ as the onset of crystallization in the outer layers that will later constitute the solid crust. Within our simplified thermal structure (isentropic interior plus an isothermal outer atmosphere), the matching point is the neutrinosphere density $\rho_\nu$, and the earliest crystallization in the envelope is therefore captured by the condition $T_\nu(t)=T_c(\rho_\nu)$. We emphasize that this marks the first appearance of a solid phase near the surface, not the completion of a fully developed crust; the crystallized region is expected to expand with time as the star continues to cool.

The critical temperature for crystallization is obtained by requiring the Coulomb interaction energy of the ions to exceed the thermal energy \citep{shap83},
\begin{align}
    k_BT_c=\frac{Z^2e^2}{\Gamma r_i},
\label{eq:kBTc}
\end{align}
where $Z$ is the characteristic ionic charge, $e$ is the elementary charge, and $\Gamma$ is the Coulomb coupling parameter at melting. The ion-sphere radius $r_i$ is defined by
\begin{align}
    \frac{4\pi}{3}n_i r_i^3=1,
\label{eq:ri_def}
\end{align}
where $n_i$ is the number density of heavy nuclei. For $\Gamma\ll 1$, ions behave approximately as an ideal Maxwell--Boltzmann gas, whereas for $\Gamma \gg 1$, Coulomb interactions dominate and the ions crystallize to minimize the Coulomb energy. Since $n_i=\rho x_a/(Am_{\rm N})=\rho(Z/A)x_a/(Zm_{\rm N})$, where $x_a$ is the mass fraction of heavy nuclei and $A$ is the typical mass number of heavy nuclei, we can explicitly give the critical temperature as\footnote{In \citet{suwa14}, $Y_e$ was used to characterize the heavy nuclei, whereas in this paper we use $Z/A$ to explicitly indicate that it represents the fraction of protons. In addition, the temperature calculation in \citet{suwa14} contained an error, which we have corrected here.}
\begin{align}
T_{c}(\rho)\approx& \frac{Z^{2} e^{2}}{\Gamma k_{B}}\left\{\frac{4\pi}{3}\frac{\rho (Z/A)x_a}{Zm_{\rm N}}\right\}^{{1/3}} %\nonumber\\
=
%& 
8.7\times 10^{9}~\mathrm{K}
\left(\frac{\Gamma}{175}\right)^{-1}
\left(\frac{\rho}{10^{14}~\mathrm{g~cm}^{-3}}\right)^{1/3}
\left(\frac{x_a}{0.3}\right)^{1/3}
\left(\frac{Z}{40}\right)^{5/3}.
\label{eq:Tc}
\end{align}
When deriving the final line, we assume $Z/A=0.1$, which was obtained, for instance, from \citet{1971NuPhA.175..225B}. The typical value of $Z$ is also based on that paper. Equation~(\ref{eq:Tc}) is used here as a fiducial one-component plasma (OCP) estimate and should not be interpreted as a unique microphysical prediction of the melting curve. Recent finite-temperature crust studies indicate that both the crystallization temperature and the composition can differ from simple OCP expectations, depending on the nuclear functional, shell effects, and multi-component treatment \citep{2020AandA...633A.149F,2020AandA...635A..84C,2020AandA...640A..77C,2023AandA...672A.160D,2023AandA...677A.174D}. In this sense, Eq.~(\ref{eq:Tc}) should be regarded as a practical upper-side fiducial scaling for the present analytic estimate. The parameters $Z$, $x_a$, and $Z/A$ are treated here as effective constants, although they can vary with density and temperature in realistic crust calculations.

\begin{figure}[tbp]
    \begin{center} 
    \includegraphics[width=0.6\textwidth]{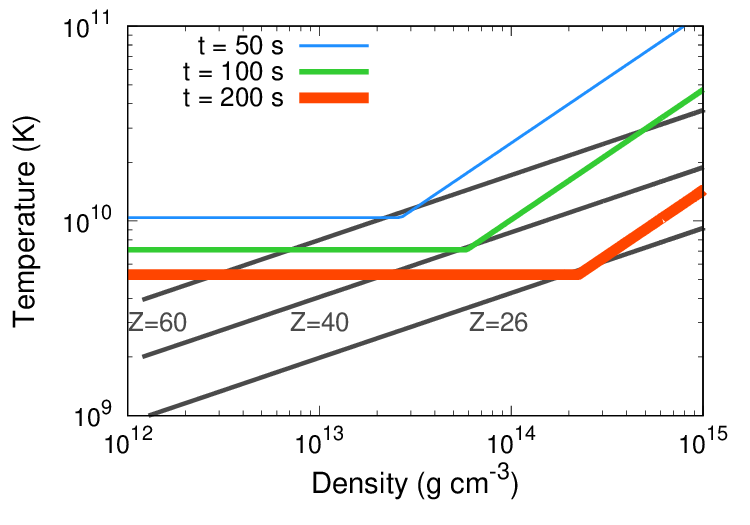}
    \end{center}
    \caption{ Temperature--density trajectories during the late PNS cooling phase and the Coulomb crystallization threshold. Solid curves show the model temperature profile $T(\rho)$ at $t=50$, 100, and 200~s for the fiducial parameters $M_{\rm PNS}=1.4\,M_\odot$, $R_{\rm PNS}=12$~km, $E_{\rm tot}=10^{53}$~erg, and $g\beta=1.6$. For each epoch, the interior follows the isentropic relation (Eq.~\ref{eq:T_profile}), while the outer atmosphere is approximated as isothermal at $T_\nu$; the neutrinosphere density $\rho_\nu$ is marked on each profile. Dashed curves give the crystallization temperature $T_c(\rho)$ (Eq.~\ref{eq:Tc}) for $Z=$26, 40, and 60 (with $\Gamma=175$, $x_a=0.3$, and $Z/A=0.1$). Crust formation sets in when $T_\nu(t)$ first drops below $T_c(\rho_\nu)$.
{Alt text: Log--log plot of temperature ($10^{9}$--$10^{11}\,\mathrm{K}$) versus density ($10^{12}$--$10^{15}\,\mathrm{g\,cm^{-3}}$) showing three piecewise curves for $t=50$ s (blue), $t=100$ s (green), and $t=200$ s (orange), overlaid with gray diagonal reference lines labeled $Z=60$, $Z=40$, and $Z=26$.}}
    \label{fig:f1}
\end{figure}
% See the instraction below for "Alt text"
% https://academic.oup.com/pasj/pages/General_Instructions#Figures%20and%20Illustrations

Figure~\ref{fig:f1} shows the temperature evolution in the $\rho$--$T$ plane. Here, we show the cases with $t=$50, 100, and 200 s after the formation of PNSs. The temperature touches the critical line with $Z=40$ at $t=100$ s at the boundary between the core and envelope. It can be seen that the crust formation condition is satisfied at $\rho=\rho_\nu$. The corresponding critical temperature is 
\begin{align}
    T_{c}(\rho_\nu)\approx& 6.5\times 10^{9}~\mathrm{K}
    \left(\frac{\Gamma}{175}\right)^{-1}
    \left(\frac{x_a}{0.3}\right)^{1/3}
    \left(\frac{Z}{40}\right)^{5/3}
    \left(\frac{M_{\rm PNS}}{1.4\,M_\odot}\right)^{1/10}
    \left(\frac{R_{\rm PNS}}{10\,\mathrm{km}}\right)^{-2/5}
    \left(\frac{g\beta}{3}\right)^{-1/10}
    \left(\frac{s}{1\,k_{B}/\mathrm{baryon}}\right)^{-1/5}.
\label{eq:Tc_rhonu}
\end{align}
With $T_\nu=T_c(\rho_\nu)$, we can calculate the entropy at the crust formation as
\begin{align}
    s_{\rm crust}=&0.12\,k_B/\mathrm{baryon}
    \left(\frac{M_{\rm PNS}}{1.4\,M_\odot}\right)^{-1/8}
    \left(\frac{R_{\rm PNS}}{10\,\mathrm{km}}\right)^{1/2}
    \left(\frac{g\beta}{3}\right)^{1/8}
    \left(\frac{\Gamma}{175}\right)^{-5/4}
    \left(\frac{x_a}{0.3}\right)^{5/12}
    \left(\frac{Z}{40}\right)^{25/12}.
\label{eq:scrust}
\end{align}
The crust formation time can be estimated by solving $s(t)=s_{\rm crust}$ as
\begin{align}
    t_{\rm crust}
    &= t_{\ast}-t_0,\\
    t_{\ast}
    &\simeq 410\,\mathrm{s}
    \left(\frac{M_{\rm PNS}}{1.4\,M_\odot}\right)^{11/12}
    \left(\frac{R_{\rm PNS}}{10\,\mathrm{km}}\right)^{-1}
    \left(\frac{g\beta}{3}\right)^{3/4}
    \left(\frac{\Gamma}{175}\right)^{1/2}
    \left(\frac{x_a}{0.3}\right)^{-1/6}
    \left(\frac{Z}{40}\right)^{-5/6},\\
    t_0
    &= 210\,\mathrm{s}
    \left(\frac{M_{\rm PNS}}{1.4\,M_\odot}\right)^{6/5}
    \left(\frac{R_{\rm PNS}}{10\,\mathrm{km}}\right)^{-6/5}
    \left(\frac{g\beta}{3}\right)^{4/5}
    \left(\frac{E_{\rm tot}}{10^{52}\,\mathrm{erg}}\right)^{-1/5}.
\label{eq:t0}
\end{align}
Using the model parameters shown in Table~1 of \citet{suwa21} (with $g\beta=g\,\beta_2$ and $E_{\rm tot}=E_{\rm tot,2}$), we obtain representative crust-formation times of $t_{\rm crust}\simeq 100$--$440\,\mathrm{s}$ for the models listed there (assuming $\Gamma=175$, $x_a=0.3$, and $Z=40$). 

In summary, Eqs.\ (\ref{eq:scrust})--(\ref{eq:t0}) provide a closed analytic estimate of the onset of crust formation during the late PNS cooling phase. The key point is that the neutrinosphere temperature decreases with time as $T_\nu\propto s^{3/5}$, while the crystallization threshold increases toward lower entropy as $T_c(\rho_\nu)\propto s^{-1/5}$. Equating these two temperatures yields an explicit entropy at crystallization, $s_{\rm crust}$, and substituting it into the analytic cooling solution gives the crust-formation time $t_{\rm crust}$. These expressions make the parameter dependence explicit: larger $M_{\rm PNS}$ and smaller $R_{\rm PNS}$ generally delay crystallization through their impact on the diffusion timescale, while higher $Z$ (and larger $x_a$) lowers $t_{\rm crust}$ by raising the melting temperature. In Section~\ref{sec:summary}, we discuss the astrophysical implications of this timescale for the early evolution of newborn neutron stars.

\section{Summary and discussion}\label{sec:summary}

In this paper, we presented a simple analytic estimate for the epoch at which a solid crust first forms in a newly born neutron star during the late protoneutron star (PNS) cooling phase. Our approach combines an analytic, diffusion-based description of the neutrino luminosity and entropy evolution (Section~\ref{sec:cooling}) with the Coulomb crystallization criterion for heavy nuclei (Section~\ref{sec:crust}). By comparing the time-dependent thermal trajectory at the neutrinosphere, $T_\nu(t)$, with the crystallization threshold evaluated at the neutrinosphere density, $T_{\rm c}(\rho_\nu)$, we derived closed expressions for the entropy at crystallization, $s_{\rm crust}$, and the corresponding onset time, $t_{\rm crust}$ (Eqs.~(\ref{eq:scrust})--(\ref{eq:t0})).

For canonical microphysics ($\Gamma=175$, $x_a=0.3$, $Z=40$) and representative PNS parameters, we find $t_{\rm crust}\sim 100$--$500\ {\rm s}$ as a typical estimate. 
The analytic formulae also make the parameter dependence explicit: larger $M_{\rm PNS}$ and/or smaller $R_{\rm PNS}$ generally delay crystallization through their impact on the diffusion timescale, whereas a larger ionic charge and heavy-nuclei mass fraction tend to accelerate crystallization by raising the melting temperature.
We stress that our estimate targets the \emph{onset} of crystallization in the outer layers around the neutrinosphere; the solid phase is expected to develop progressively with radius and depth as the star continues to cool.

Within this limited scope, crust onset may provide a useful characteristic transition time for later magnetic-field evolution, because the outer layers change from fluid-like to solid and highly conducting. The present paper, however, does not attempt a quantitative model of magnetic-field relaxation or field-configuration outcomes.

Several caveats should be kept in mind. The analytic cooling expressions are extrapolated beyond their main calibration interval, and by $t\sim100\,\mathrm{s}$ the PNS interior is already becoming increasingly semitransparent to neutrinos, so a purely diffusive description is no longer everywhere accurate. In addition, the composition parameters $Z$, $x_a$, and $Z/A$ are treated as effective constants despite their likely spatial and temporal variations \citep{2025PTEP.2025l3E01N}, while the melting prescription is based on a fiducial OCP Coulomb criterion. A quantitatively predictive treatment of crust formation therefore requires multidimensional neutrino-radiation hydrodynamics together with improved finite-temperature crust microphysics.

%\section{Section3}\label{sec:3}
%
%Figure captions should always begin with a declarative title, followed by a brief description (see figure \ref{fig:sample}). 
%In each table, a declarative title should be given in \verb/\tbl{ }/ or \verb/\caption{ }/. 
%Any notes applying to the table and specific parts appear immediately below the table with symbols.   

%%%%%%%%%%%%%%%%%%%%%%%%%%%%%%%%%%%%%%%

\begin{ack}
This work is supported by JSPS KAKENHI grant Nos. JP24H02245, JP24K00668, JP24K07021, and JP25K01035.
\end{ack}


\begin{thebibliography}{30}
\expandafter\ifx\csname natexlab\endcsname\relax\def\natexlab#1{#1}\fi

\bibitem[{{Baym} {et~al.}(1971{\natexlab{a}}){Baym}, {Bethe}, \&
  {Pethick}}]{1971NuPhA.175..225B}
{Baym}, G., {Bethe}, H.~A., \& {Pethick}, C.~J. 1971{\natexlab{a}}, Nuclear
  Physics A, 175, 225

\bibitem[{{Baym} {et~al.}(1971{\natexlab{b}}){Baym}, {Pethick}, \&
  {Sutherland}}]{1971ApJ...170..299B}
{Baym}, G., {Pethick}, C.~J., \& {Sutherland}, P. 1971{\natexlab{b}}, The
  Astrophysical Journal, 170, 299

\bibitem[{{Bethe}(1990)}]{beth90}
{Bethe}, H.~A. 1990, Reviews of Modern Physics, 62, 801

\bibitem[{{Burrows} \& {Lattimer}(1986)}]{1986ApJ...307..178B}
{Burrows}, A., \& {Lattimer}, J.~M. 1986, The Astrophysical Journal, 307, 178

\bibitem[{{Carreau} {et~al.}(2020{\natexlab{a}}){Carreau}, {Fantina}, \&
  {Gulminelli}}]{2020AandA...640A..77C}
{Carreau}, T., {Fantina}, A.~F., \& {Gulminelli}, F. 2020{\natexlab{a}},
  Astronomy and Astrophysics, 640, A77

\bibitem[{{Carreau} {et~al.}(2020{\natexlab{b}}){Carreau}, {Gulminelli},
  {Chamel}, {Fantina}, \& {Pearson}}]{2020AandA...635A..84C}
{Carreau}, T., {Gulminelli}, F., {Chamel}, N., {Fantina}, A.~F., \& {Pearson},
  J.~M. 2020{\natexlab{b}}, Astronomy and Astrophysics, 635, A84

\bibitem[{{Chamel} \& {Haensel}(2008)}]{2008LRR....11...10C}
{Chamel}, N., \& {Haensel}, P. 2008, Living Reviews in Relativity, 11, 10

\bibitem[{{Dinh Thi} {et~al.}(2023{\natexlab{a}}){Dinh Thi}, {Fantina}, \&
  {Gulminelli}}]{2023AandA...677A.174D}
{Dinh Thi}, H., {Fantina}, A.~F., \& {Gulminelli}, F. 2023{\natexlab{a}},
  Astronomy and Astrophysics, 677, A174

\bibitem[{{Dinh Thi} {et~al.}(2023{\natexlab{b}}){Dinh Thi}, {Fantina}, \&
  {Gulminelli}}]{2023AandA...672A.160D}
---. 2023{\natexlab{b}}, Astronomy and Astrophysics, 672, A160

\bibitem[{{Fantina} {et~al.}(2020){Fantina}, {De Ridder}, {Chamel}, \&
  {Gulminelli}}]{2020AandA...633A.149F}
{Fantina}, A.~F., {De Ridder}, S., {Chamel}, N., \& {Gulminelli}, F. 2020,
  Astronomy and Astrophysics, 633, A149

\bibitem[{{Fischer} {et~al.}(2010){Fischer}, {Whitehouse}, {Mezzacappa},
  {Thielemann}, \& {Liebend{\"o}rfer}}]{2010AandA...517A..80F}
{Fischer}, T., {Whitehouse}, S.~C., {Mezzacappa}, A., {Thielemann}, F.-K., \&
  {Liebend{\"o}rfer}, M. 2010, Astronomy and Astrophysics, 517, A80

\bibitem[{{Haensel} {et~al.}(2007){Haensel}, {Potekhin}, \&
  {Yakovlev}}]{2007ASSL..326.....H}
{Haensel}, P., {Potekhin}, A.~Y., \& {Yakovlev}, D.~G. 2007, Astrophysics and
  Space Science Library, Vol. 326, {Neutron Stars 1: Equation of State and
  Structure} (Springer)

\bibitem[{{H{\"u}depohl} {et~al.}(2010){H{\"u}depohl}, {M{\"u}ller}, {Janka},
  {Marek}, \& {Raffelt}}]{2010PhRvL.104y1101H}
{H{\"u}depohl}, L., {M{\"u}ller}, B., {Janka}, H.-T., {Marek}, A., \&
  {Raffelt}, G.~G. 2010, Physical Review Letters, 104, 251101

\bibitem[{{Janka}(2012)}]{2012ARNPS..62..407J}
{Janka}, H.-T. 2012, Annual Review of Nuclear and Particle Science, 62, 407

\bibitem[{{Janka}(2025)}]{2025ARNPS..75..425J}
---. 2025, Annual Review of Nuclear and Particle Science, 75, 425

\bibitem[{{Mirizzi} {et~al.}(2016){Mirizzi}, {Tamborra}, {Janka}, {Saviano},
  {Scholberg}, {Bollig}, {H{\"u}depohl}, \&
  {Chakraborty}}]{2016NCimR..39....1M}
{Mirizzi}, A., {Tamborra}, I., {Janka}, H.-T., {Saviano}, N., {Scholberg}, K.,
  {Bollig}, R., {H{\"u}depohl}, L., \& {Chakraborty}, S. 2016, Rivista del
  Nuovo Cimento, 39, 1

\bibitem[{{Nagakura} {et~al.}(2020){Nagakura}, {Burrows}, {Radice}, \&
  {Vartanyan}}]{2020MNRAS.492.5764N}
{Nagakura}, H., {Burrows}, A., {Radice}, D., \& {Vartanyan}, D. 2020, Monthly
  Notices of the Royal Astronomical Society, 492, 5764

\bibitem[{{Nakazato} {et~al.}(2013){Nakazato}, {Sumiyoshi}, {Suzuki}, {Totani},
  {Umeda}, \& {Yamada}}]{2013ApJS..205....2N}
{Nakazato}, K., {Sumiyoshi}, K., {Suzuki}, H., {Totani}, T., {Umeda}, H., \&
  {Yamada}, S. 2013, The Astrophysical Journal Supplement Series, 205, 2

\bibitem[{{Nakazato} {et~al.}(2018){Nakazato}, {Suzuki}, \& {Togashi}}]{naka18}
{Nakazato}, K., {Suzuki}, H., \& {Togashi}, H. 2018, Physical Review C, 97,
  035804

\bibitem[{{Nakazato} {et~al.}(2025){Nakazato}, {Togashi}, {Sumiyoshi}, \&
  {Suzuki}}]{2025PTEP.2025l3E01N}
{Nakazato}, K., {Togashi}, H., {Sumiyoshi}, K., \& {Suzuki}, H. 2025, Progress
  of Theoretical and Experimental Physics, 2025, 123E01

\bibitem[{{Pascal} {et~al.}(2022){Pascal}, {Novak}, \&
  {Oertel}}]{2022MNRAS.511..356P}
{Pascal}, A., {Novak}, J., \& {Oertel}, M. 2022, Monthly Notices of the Royal
  Astronomical Society, 511, 356

\bibitem[{{Pons} {et~al.}(1999){Pons}, {Reddy}, {Prakash}, {Lattimer}, \&
  {Miralles}}]{1999ApJ...513..780P}
{Pons}, J.~A., {Reddy}, S., {Prakash}, M., {Lattimer}, J.~M., \& {Miralles},
  J.~A. 1999, The Astrophysical Journal, 513, 780

\bibitem[{{Pons} \& {Vigan{\`o}}(2019)}]{2019LRCA....5....3P}
{Pons}, J.~A., \& {Vigan{\`o}}, D. 2019, Living Reviews in Computational
  Astrophysics, 5, 3

\bibitem[{{Roberts} {et~al.}(2012){Roberts}, {Shen}, {Cirigliano}, {Pons},
  {Reddy}, \& {Woosley}}]{2012PhRvL.108f1103R}
{Roberts}, L.~F., {Shen}, G., {Cirigliano}, V., {Pons}, J.~A., {Reddy}, S., \&
  {Woosley}, S.~E. 2012, Physical Review Letters, 108, 061103

\bibitem[{{Rusakov} {et~al.}(2026){Rusakov}, {Burrows}, {Wang}, \&
  {Vartanyan}}]{2026arXiv260209025R}
{Rusakov}, A., {Burrows}, A.~S., {Wang}, T., \& {Vartanyan}, D. 2026, arXiv
  e-prints, arXiv:2602.09025

\bibitem[{{Shapiro} \& {Teukolsky}(1983)}]{shap83}
{Shapiro}, S.~L., \& {Teukolsky}, S.~A. 1983, {Black Holes, White Dwarfs, and
  Neutron Stars: The Physics of Compact Objects} (New York: Wiley-Interscience)

\bibitem[{{Suwa}(2014)}]{suwa14}
{Suwa}, Y. 2014, Publications of the Astronomical Society of Japan, 66, L1

\bibitem[{{Suwa} {et~al.}(2021){Suwa}, {Harada}, {Nakazato}, \&
  {Sumiyoshi}}]{suwa21}
{Suwa}, Y., {Harada}, A., {Nakazato}, K., \& {Sumiyoshi}, K. 2021, Progress of
  Theoretical and Experimental Physics, 2021, 013E01

\bibitem[{{Suwa} {et~al.}(2019){Suwa}, {Sumiyoshi}, {Nakazato}, {Takahira},
  {Koshio}, {Mori}, \& {Wendell}}]{2019ApJ...881..139S}
{Suwa}, Y., {Sumiyoshi}, K., {Nakazato}, K., {Takahira}, Y., {Koshio}, Y.,
  {Mori}, M., \& {Wendell}, R.~A. 2019, The Astrophysical Journal, 881, 139

\bibitem[{{Yamada} {et~al.}(2024){Yamada}, {Nagakura}, {Akaho}, {Harada},
  {Furusawa}, {Iwakami}, {Okawa}, {Matsufuru}, \&
  {Sumiyoshi}}]{2024PJAB..100..190Y}
{Yamada}, S., {et~al.} 2024, Proceedings of the Japan Academy, Series B, 100,
  190

\end{thebibliography}
\end{document}